\documentclass[]{elsarticle}

\usepackage{hyperref}
\usepackage{float,lscape}
\usepackage{graphicx}
\usepackage{parskip}
\usepackage{rotating}
\usepackage{amsmath}
\usepackage{longtable}
\usepackage{lscape}
\usepackage{caption}

\journal{Proceedings A of Royal Society}

\begin{document}

\begin{frontmatter}

\title{Monte Carlo model for electron degradation in Xenon gas}
\tnotetext[]{Article accepted in Proceedings A of Royal Society DOI://dx.doi.org/10.1098/rspa.2015.0727}

\author{Vrinda Mukundan\corref{cor1}}
\ead{vrinda\_mukundan@vssc.gov.in}

\author{ Anil Bhardwaj\corref{cor2}} 
\ead {anil\_bhardwaj@vssc.gov.in, bhardwaj\_spl@yahoo.com}

\address{Space Physics Laboratory,
Vikram Sarabhai Space Centre,
Trivandrum~695022, India}

\begin{abstract}
 We have developed a Monte Carlo model for studying the local degradation of electrons in the energy range 9-10000 eV
 in xenon gas. Analytically fitted form of  electron impact cross sections for elastic and various inelastic processes 
 are fed as input data to the model. Two dimensional numerical yield spectrum, which gives information on the number of 
 energy loss events occurring in a particular energy interval, is obtained as output of the model. Numerical yield spectrum 
 is fitted analytically, thus obtaining analytical yield spectrum. The analytical yield spectrum can be used to calculate 
 electron fluxes, which can be further employed for the calculation of volume production rates. Using yield spectrum, 
 mean energy per ion pair and efficiencies of inelastic processes are calculated. The value for mean energy per ion pair for Xe is 22 eV 
 at 10 keV. Ionization dominates for incident energies greater than 50 eV and is found to have an efficiency of 
 $\sim$65\% at 10 keV. The efficiency for the excitation process is  $\sim$30\% at 10 keV.
\end{abstract}


\end{frontmatter}


\section{Introduction}
When electrons interact with atoms or molecules, energy of the incident electron is lost through 
inelastic collisions with the target. The target atom can undergo ionization or excitation. Secondary 
electrons released during ionization process will trigger further collisions. These electrons lose energy 
via further collisions in the gas. 
This kind of electron energy degradation process is of great importance in understanding phenomena, like electron beam propagation in the 
atmosphere, population inversion process in a large group of gas lasers, optical emissions occurring in the upper 
atmosphere, like aurora and airglow which happen due to the precipitation of high energy electrons, etc 
\cite{2006Trajmar, 1986Sorokin, 2009Campbell, 2000Bhardwaj}.

Xenon is a widely studied atom due to its inert nature and simple ground state atomic structure. 
The inert nature of xenon  makes it possible to study the pristine environment of the early solar system 
whose traces have been removed by other reactive elements. Noble gases are valuable probes of extraterrestrial 
environments. Their abundances as well as isotopic compositions are indicators of various processes such as stellar 
events prior to solar system formation and radioactive decay. Hence their study is vital to understand the sequence 
of events that led to the formation and subsequent evolution  of the solar system. There are various missions that studied
xenon in planetary atmospheres. For example, Galileo mass spectrometer reported a relative abundance of 2.6$\pm$0.5 times 
solar ratios in Jovian atmosphere \cite{2000Mahaffy}. Pioneer Venus Sounder Probe Neutral Mass Spectrometer 
measured an upper limit for xenon in Venusian atmosphere as 120 ppb \cite{1981Donahue}. In Earth's atmosphere 
xenon is a trace gas, having a concentration of 87 ppb \cite{1980Fleagle}.

Electron collision with Xe has a wide range of application. 
In x-ray and gamma ray detectors, Xe gas counters have been commonly
used. In both these detectors, interaction of radiation with Xe atoms results in the production of electrons. Among the various
methods of amplifying the obtained skimp signal, a commonly used approach is electron acceleration in which the accelerated 
electrons excite the xenon atoms through inelastic collisions leading to 
the production of secondary scintillation, i.e. the emission of detectable light in the vacuum ultraviolet region\cite{2000Knoll}.  
The same technique is used in flash lamps which produce high intensity white light for 
a short duration. 
The excited atoms, created by electron-atom collisions, can de-excite by emitting in different wavelengths whose combined effect will give 
the appearance of white light emission.
Electron bombardment technique with xenon is also used in ion thrusters which is an 
evolving technology in the field of rocket propulsion.  Here, electrically accelerated Xe ions, created using electron impact, 
are emitted at high speed as exhaust and this will push the spacecraft forward. Efficiency of this kind 
of ion thrusters are found to be higher than that of conventional chemical propulsion methods and has been used in NASA's missions 
Deep Space-1 \cite{1999Rayman} and  DAWN \cite{2006Rayman}. The utmost use of xenon gas for such purposes will be possible only through a 
thorough understanding of properties associated with microscopic collision processes. 

Monte Carlo simulation
is a commonly used method for studying particle energy degradation process
\cite{2015Bhardwaj, Bhardwaj09, Bhardwaj99a, 1980Singhal, 1977Green, 1971Cicerone}. The electron collision process in 
xenon was studied earlier by Date et al. \cite{2003Date} using Monte Carlo method.
Rachinhas et al. \cite{1999Rachinhas} studied the absorption of electrons with energies $\leq$ 200 keV in xenon using 
Monte Carlo simulation technique. They calculated mean energy per ion pair (w-value) and Fano factor in the energy range 20-200 keV
and studied the influence of electric field on the results. Absorption of x-ray photons and the drift of 
resulting electrons under the influence of an applied electric field in xenon was studied by Dias et al. \cite{1993Dias}, by 
using Monte Carlo technique. A 3D Monte Carlo method was used by Santos et al. \cite{1994Santos} to study the drift of electrons
in xenon and to calculate various physical properties such as electroluminescence, drift velocities etc.

We have developed a Monte Carlo model for the local degradation of electrons with energy in the range 9-10000 eV
in neutral xenon gas to understand how the energy of the incident electron will be distributed among various loss channels like 
ionization and excitation while it is making collisions with neutral xenon atoms. 
Using our Monte Carlo model, we have calculated numerical yield spectra which is a basic distribution function that contain 
information regarding the degradation process and can be used to calculate the yield of any excited or ionized states. 
The numerical yield spectra is then fitted analytically, thus obtaining analytical yield spectra.
For practical purposes, the use of analytical yield spectrum simplify the application by substantially
reducing the computational time. We have also obtained secondary 
electron energy distribution, the efficiency of
excitation and ionization processes, as well as the mean energy expended for ion pair production. 

\section{Monte Carlo model}
In the Monte Carlo simulation, the energy loss 
process of the electron is treated in a discrete manner. In carrying out the degradation by means of discrete steps, the electron is 
followed as it undergoes successive collisions. To accomplish the energy degradation in a convenient way, the energy range 
from the incident energy to cut off energy is divided into a number of equally spaced bins. Whenever an electron makes an inelastic collision, 
the collision event is recorded in the corresponding energy bin. This process is continued and the particle, its secondaries, tertiaries, etc. 
are followed until their energy falls below an assigned cut off value.  

The energy bin size is taken as 1 eV throughout the energy range.
To make an inelastic collision with a xenon atom, an electron should have a minimum energy of 8.315 eV as it is the 
lowest threshold of all the inelastic processes. We have set the cut off as 9 eV since
we have 1 eV energy bins in the model.

\begin{figure}
\noindent\includegraphics[width=1.0\linewidth]{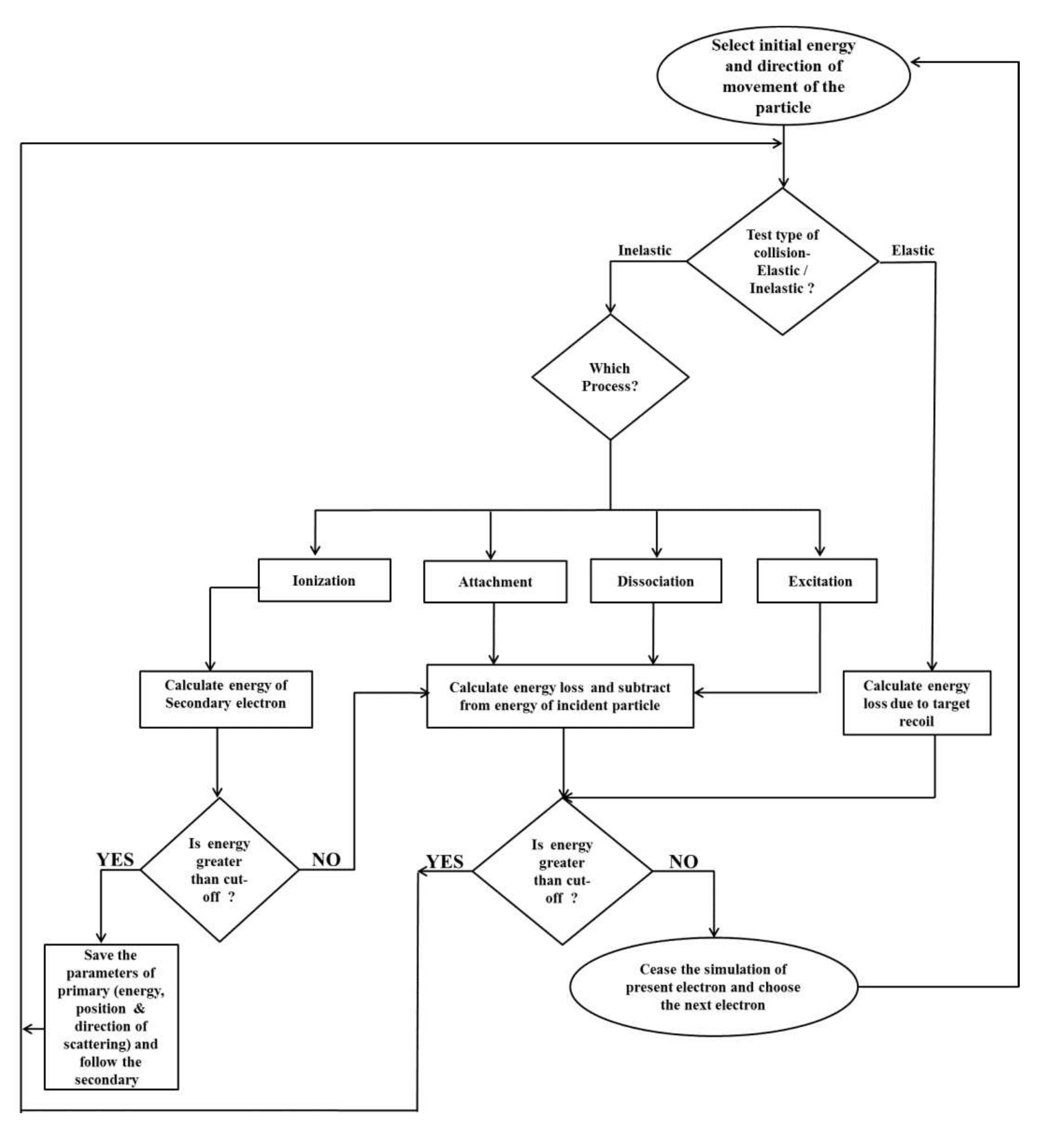}
\caption{Flowchart showing Monte Carlo simulation for electron degradation. Figure shows 
flow only up to secondary electrons; subsequent electrons (tertiary, quaternary etc.) are also followed in the similar 
manner in the simulation.}
\label{flowchart}
\end{figure}

Figure \ref{flowchart} shows the flow diagram for our Monte Carlo simulation for electron degradation.
The simulation starts by fixing the energy of the incident electron. The direction of movement of electron ($\theta$, $\phi$) is assumed 
to be isotropic. The distance that the electron has to travel before the collision is calculated as 

\begin{equation}
S = -\log(1-R)/N\sigma_T,        
\end{equation}
where R is a random number, N is the number density of the gas (equal to 10$^{10}$ cm$^{-3}$) and 
$\sigma_T$ is the total scattering cross section (elastic + inelastic). Next we decide on the type of collision. 
The probabilities of the elastic and inelastic events, P$_{el} $($\sigma_{el}$/ $\sigma_T$) and P$_{in}$ ($\sigma_{in}$/ $\sigma_T$), 
 where $\sigma_{el}$ and $\sigma_{in}$ are elastic and inelastic cross sections, are  
calculated and compared with new random number R$_4$. Elastic collision occurs if P$_{el}$ $\geq$ R$_{4}$. 
The energy loss in elastic collisions $\Delta$E  due to target recoil is calculated  as 

\begin{equation}
\bigtriangleup E=\frac{m^2v^2}{m+M}-\frac{m^2vV_1\cos\delta}{m+M},
\end{equation}  
where
$$
V_1=v\left[\frac{m\cos\delta}{m+M}+\frac{[M^2+m^2(\cos\delta-1)]^{1/2}}
{m+M}\right].
$$
Here $\delta$ is the scattering angle in the laboratory frame, $v$ and
$m$ are, respectively, the velocity and mass of the incident electron, and $M$
is the mass of the target particle. The scattering angle $\delta$ is determined 
by using differential elastic cross sections which are fed numerically into the model. 
The energy lost in the collision is then subtracted from the incident electron energy. After the collision, the 
deflection angle relative to the direction ($\theta$,$\phi$) is obtained by

$$
\cos\theta^{''}=\cos\theta\cos\theta^{'}-\sin\theta\sin\theta^{'}
\cos\phi^{'} ,                    
$$
\begin{equation}
\cos\phi^{''}=(\cos\theta\cos\phi\sin\theta^{'}
            \sin\phi^{'}-\sin\phi\sin\theta^{'}\sin\phi^{'}
           +\sin\theta\cos\phi\cos\theta^{'})/\sin\theta^{''}, \label{direction}
\end{equation}
$$
\sin\phi^{''}=(\cos\theta\cos\phi\sin\theta^{'}\cos\phi^{'}
            -\cos\phi\sin\theta^{'}\sin\phi^{'}                                 
            +\sin\theta\sin\phi\cos\theta^{'})/\sin\theta^{''}.
$$
Here $\theta^{'}$, $\phi^{'}$ are the scattering angles.

If an inelastic collision occurs, the collision event is recorded in the appropriate 
energy bin corresponding to the energy of the particle. It is further decided whether it is an excitation or ionization event. In case of ionization event,
the energy of secondary electron has to be calculated as it can also initiate further inelastic collisions, provided 
it has sufficient energy. The secondary electron energy is calculated as \cite{1972Green}

\begin{equation}
  T=\frac{\Gamma_S\ E_v}{E_v+\Gamma_B}[\tan(RK_1+(R-1)K_2)]+T_S
     -\left[\frac{T_A}{E_v+T_B}\right],          \label{2e-eqn}
\end{equation}
where
$$
K_1 = \tan^{-1}\left\{\left[\frac{(E_v-I)}{2}-T_S
    +\frac{T_A}{(E_v+T_B)}\right]
    \bigg/\frac{\Gamma_S\ E_v}{(E_v+\Gamma_B)}\right\},
$$
$$
K_2 = \tan^{-1}\left\{\left[T_S
     -\frac{T_A}{(E_v+T_B)}\right]
     \bigg/\frac{\Gamma_S\ E_v}{(E_v+\Gamma_B)}\right\}.
$$
Here $E_v$ is the incident electron energy; $\Gamma_S$, $\Gamma_A$, $T_A$, 
$T_B$, and $T_S$ are
the fitting parameters, and $I$ is the ionization threshold. 
We have used the fitting parameters of Green and Sawada \cite{1972Green}. If this energy is 
greater than that of the cut off energy (9 eV)  then the secondary electron has to be followed. 
In order to follow the secondary electron, the parameters of the primary electron, i.e. the energy 
remaining in the primary, its position and direction of movement are first saved in suitable variables. Secondary 
electron is then followed in the same method as the primary electron. Once the energy of the secondary 
is completely degraded, the saved parameters of the primary electron are retrieved and its degradation is continued. 
Similarly, tertiary, quaternary, etc., electrons are followed in the simulation. 

For all inelastic collisions, the collision event is recorded in the corresponding energy bin so that the information 
on the total number of collisions that occur in each energy bin can be obtained, once the simulation is complete. This is used for 
calculating the yield spectrum and is described in detail in Section 4a.
The number of secondary, tertiary, quaternary, etc., electrons produced during ionization events are also stored in the corresponding energy bins 
which is used to determine their energy distribution (See Section 4c). The angle and direction of movement 
of electron after each ionization and excitation event are calculated using differential elastic cross sections as described in equation \ref{direction}.
After each inelastic collision, appropriate energy is subtracted from the particle energy.
If the remaining energy is higher than cutoff energy, it is again followed in the simulation. 
The simulation is made for a monoenergetic beam of 
10$^6$ electrons; each and every electron is followed in a collision-by-collision manner until its energy falls below 9 eV.  

Modeling the electron energy degradation primarily requires a set of electron impact excitation and ionization cross sections for the atom.
These cross sections are essential for electron energy deposition schemes and are presented below in detail.

\section{Cross Sections}
\subsection{Total Elastic cross sections}

Total elastic scattering cross sections for Xe have been measured or calculated
by many authors, like Mayol and Salvat\cite{1997Mayol}, Gibson et al. \cite{1998Gibson}, 
Adibzadeh  and Theodosiou \cite{2005Adibzadeh},
Vinodkumar et al. \cite{2007Vinodkumar} and McEachran and Stauffer \cite{2014Stauffer}. In the present model, we have used the 
analytically fitted theoretical cross sections of McEachran and Stauffer \cite{2014Stauffer} which are 
calculated using a relativistic optical potential method. These cross section are in 
good agreement with the measured values of Mayol and Salvat \cite{1997Mayol} and also with the theoretical values of 
Adibzadeh and Theodosiou \cite{2005Adibzadeh} and Vinodkumar et al. \cite{2007Vinodkumar} in the energy 
range 50-1000 eV. However, at energies between 15 eV and 50 eV, cross sections of Vinodkumar et al. \cite{2007Vinodkumar} 
is lower with a maximum deviation of 50\% at 30 eV. Calculations by McEachran and Stauffer \cite{2014Stauffer} is
higher than the cross sections of Gibson et al. \cite{1998Gibson} and Adibzadeh  and Theodosiou \cite{2005Adibzadeh} at 1-10 eV. 
The maximum deviation (25\%) is found at 6 eV. We have extended the analytical fit of McEachran and Stauffer \cite{2014Stauffer} 
to 10 keV to calculate the cross section at higher energies. This extension is valid as it agrees with the cross sections calculated by Gracia et al. \cite{2002Gracia}
using a scattering potential method. 

\subsection{Differential elastic cross sections}
The direction in which electron is scattered after each collision is calculated using differential elastic scattering cross sections (DCS).
For the present work, DCS of Adibzadeh  and Theodosiou \cite{2005Adibzadeh} is used, in which values for the energy 
range 1-1000 eV are given for a finer energy grid (1 eV). These values are in good agreement with 
the DCS values of McEachran and Stauffer \cite{1984McEachran} and Sienkiewicz and Baylis \cite{1989Sienkiewicz}.
For energies greater than 1000 eV, linearly extrapolated values of differential cross sections are used, as measurements are not available. DCS values at few energies 
are shown in Table \ref{tab-EDCS}.

\subsection{Ionization cross sections}

Both single and multiple ionization cross sections of xenon have been measured by 
Schram \cite{1966Schram}, Nagy et al. \cite{1980Nagy}, Stephan and Mark,\cite{1984Stephan}, 
Wetzel et al. \cite{1987Wetzel}, Lebius et al. \cite{1989Lebius}, 
 Krishnakumar and Srivastava \cite{1988Krishnakumar}, Almeida \cite{2002Almeida}, and Rejoub et al. \cite{2002Rejoub}.
Up to 1000 eV, we have used the recent measurements of  Rejoub et al. \cite{2002Rejoub} which are in 
good agreement with the work of Rapp and Englander-Golden \cite{1965Rapp}, Schram \cite{1966Schram}, 
Nagy et al. \cite{1980Nagy} and Stephan and Mark \cite{1984Stephan}. At energies greater than 1000 eV, 
measurements of Schram \cite{1966Schram} have been used as it is the only available 
measurements at higher energies. These cross sections are fitted using empirical formula of Krishnakumar and Srivastava \cite{1988Krishnakumar};

\begin{equation}
   \sigma_{p}(E) = \frac{1}{IE}[ A ln\frac{E}{I} + \sum_{i=1}^N B_{i}(1-\frac{I}{E})^{i} ]
  \end{equation}

where A and B$_i$ are fitting coefficients, I is the ionization threshold, E is the 
electron energy and i is the number of terms N required to fit the data. Fitting parameters had to be adjusted
as the cross sections of Krishnakumar and Srivastava \cite{1988Krishnakumar} are higher than that measured by  
 Rejoub et al. \cite{2002Rejoub} and Schram \cite{1966Schram} by about 20\% at the maximum. Fitting parameters used in the present study are given in 
Table \ref{ion-xs-fitting}. 

\begin{figure}[h]
\noindent\includegraphics[width=0.75\linewidth]{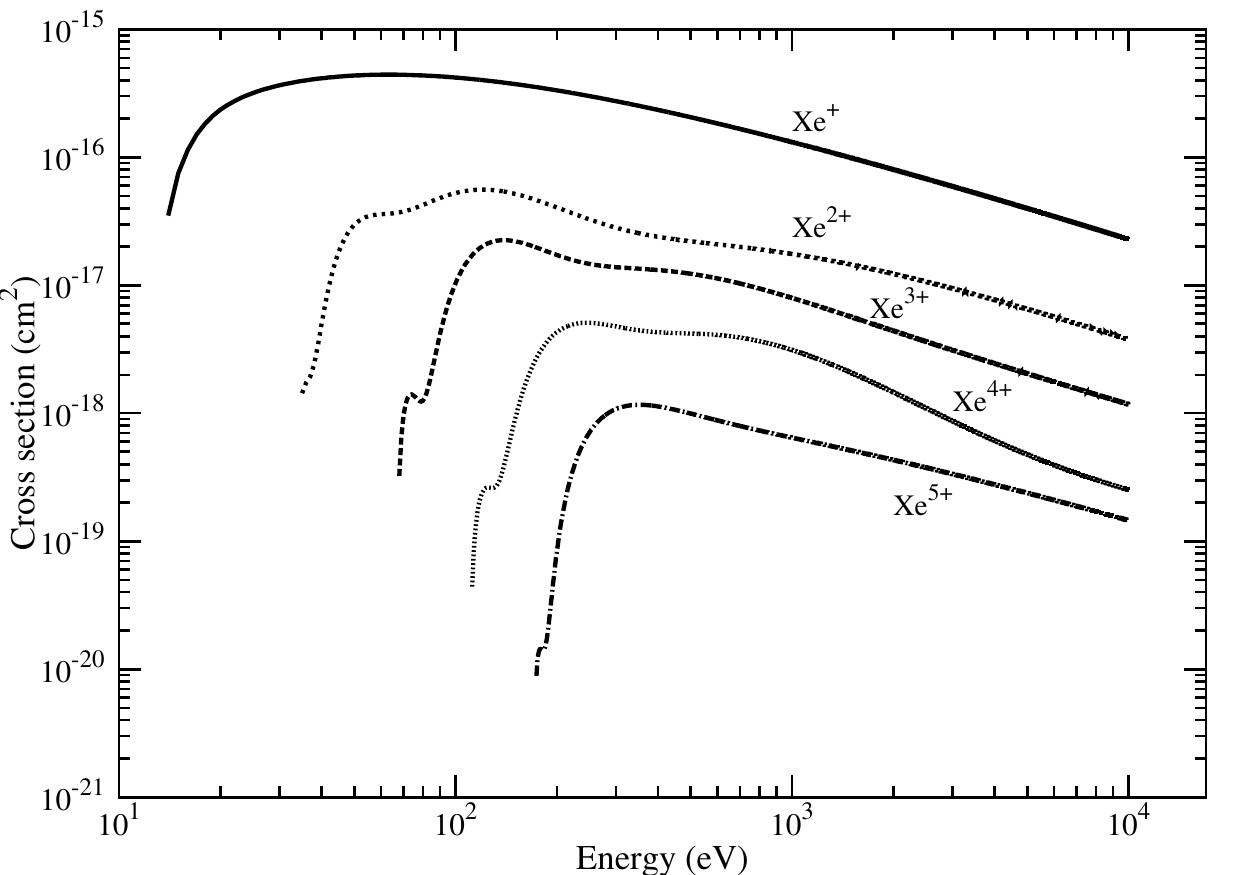}
\caption{Electron impact ionization cross sections for Xe. Cross sections up to five 
ionization states are shown here.}
\label{cross-ion}
\end{figure}

In our model we have considered only up to the fifth ionization state of xenon. Higher 
states have very low ionization cross sections and the total yield will remain more or less
the same even if they are taken into account. Ionization cross sections used in the model 
are shown in Figure \ref{cross-ion}. These partial ionization cross sections are then used 
to calculate gross ionization cross section  as 

\begin{equation}
 \sigma_{gross} = \sigma^{+} + 2\sigma^{2+} + 3\sigma^{3+} +  4\sigma^{4+} + 5\sigma^{5+} \label{gross-ion}
\end{equation}

As xenon is a gas that is capable of multiple ionization, total ionization cross section 
will be the charge weighted sum of partial ionization cross sections \cite{1965Rapp}. It is 
this gross ionization cross section that is used in the model.

\subsection{Excitation cross sections}

Xenon (Z = 54) has a ground state configuration of 5$p^6$. Electron impact excitation 
can result in configurations like 5$p^5ns$, 5$p^5np$, 5$p^5nd$ etc. Each of these excited 
configurations will be composed of different levels which occur due to the coupling between 
the core angular momentum $J_c$ and the angular momentum of the excited electron. For example, 
the 5$p^5$6$s$ configuration is composed of four levels which are represented as 1$s_2$, 1$s_3$, 
1$s_4$ and 1$s_5$ (in the decreasing order of energy) in Paschen notation with $J$ values 1, 0, 
1 and 2, respectively. Excitation cross sections for the various excitation levels of Xe are available 
in the literature. However, individual cross sections of various levels in each configuration have not been 
calculated.

Cross sections for the excitation into 5$p^5$7$p$ levels from the 
ground level as well as from the 5$p^5$6$s$ levels of xenon were measured by Jung et al. \cite{2009Jung}. 
Sharma et al. \cite{2011Sharma} theoretically calculated the cross sections for the excitation into the  
5$p^5$7$p$ levels. Excitation cross section from the ground state to the 5$p^5$6$s$ level 
 was measured by Fons and Lin \cite{1998Fons}. Puech and Mizzi \cite{1991Puech} reported cross sections
 for the 13 excited levels of xenon where the excitation cross sections for the forbidden and allowed 
 transitions were calculated separately. They made use of Born-Bethe approximation to calculate
the cross sections at high incident electron energies and a low energy modifier to extend 
the calculations down to threshold energies. These semi-empirical expressions which are valid from threshold to
relativistic energies are used in the current model. Excitation cross section for an allowed level is 
calculated as 

\begin{equation}
   \sigma_j = \frac{8\pi a_{0}^{2}R^{2}}{mc^{2}\beta^{2}}\frac{F_{oj}}{W_{j}}[ ln(\frac{\beta^2}{1-\beta^2}\frac{mc^2\beta^2}{2W_j})-\beta^2] \label{exct1}
\end{equation}

where a$_o$ is the Bohr radius, R the Rydberg constant, m the rest mass of electron and $\beta$ is the velocity of incident electron 
in units of light velocity c. W$_j$ is the excitation threshold of the j$^{th}$ level and F$_{oj}$ is the oscillator strength. For 
forbidden states, cross sections are calculated as  
\begin{equation}
  \sigma_j = \frac{8\pi a_{0}^{2}R}{mc^{2}\beta^{2}}F_j \label{exct2}
\end{equation}
where F$_j$ is a constant. To calculate cross sections at energies near threshold region, equations (\ref{exct1}) 
and (\ref{exct2}) have to be multiplied by  a low energy modifier 
\begin{equation}
 B_j = \frac{[1-(\frac{2W_j}{mc^{2}\beta^{2}})^{a_j}]^{b_j}}{(mc^{2}\beta^{2})^{c_j}}
\end{equation}
where a$_j$, b$_j$ and c$_j$ are fitting parameters. Values of these parameters for the different excitation levels 
are shown in Table \ref{tab-exct-para}. Cross sections for various excitations are added together 
to obtain total excitation cross section.

\subsection{Total cross sections}
\begin{figure}
\noindent\includegraphics[width=0.75\linewidth]{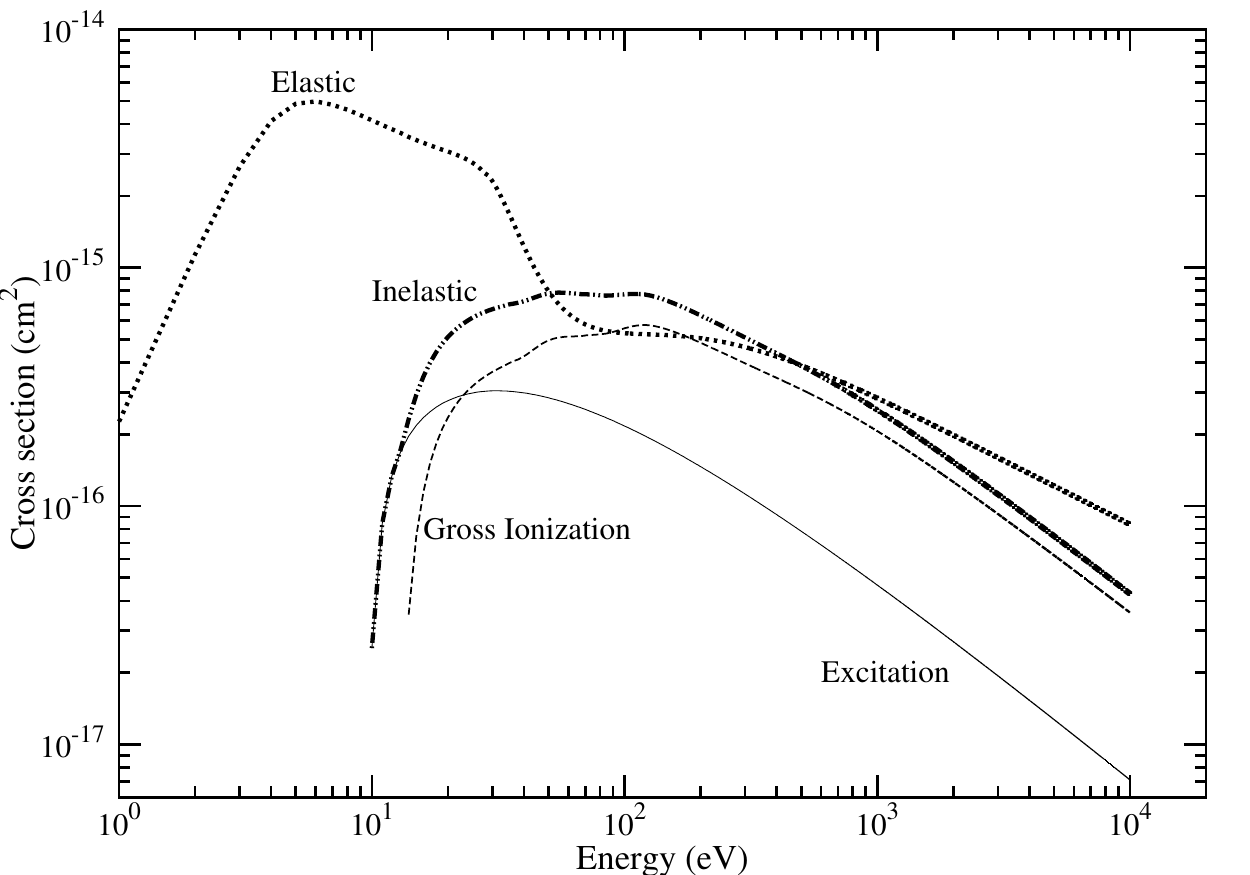}
\caption{Electron impact cross sections for various elastic and inelastic processes of Xe.}
\label{cross-all}
\end{figure}

Total inelastic cross section is calculated by adding total excitation cross section and gross ionization 
cross section. These total inelastic cross sections and elastic cross sections are added up to obtain total scattering 
cross sections. Our calculated total scattering cross sections are in good agreement with values of 
Kurokawa et al. \cite{2011kurokawa}, Zecca et al. \cite{1991Zecca} and Vinodkumar et al. \cite{2007Vinodkumar}. 
Figure \ref{cross-all} shows various cross sections that are used in our model. 

\section{Results}
\subsection{Yield Spectrum}

Yield spectrum, U(E, E${_0}$), for an incident electron energy E${_0}$ and spectral energy E, 
is defined as the number of discrete energy loss events that happened in an energy interval E and 
E+$\Delta$E. 
\begin{equation}
      U(E,E_0)=\frac{N(E)}{\bigtriangleup E}, \label{yield}
\end{equation}
where N(E) is the number of inelastic collisions and $\bigtriangleup E$ is the energy bin width which is 
1 eV in our model. This yield spectrum can be used for calculating the population (J) of any 
state j , which is the number of inelastic events of type j caused by an electron while degrading its
energy from E$_0$ to cut off as 

\begin{equation}
     J_j(E_0)=\int_{W_{th}}^{E_0} U(E,E_0)\: P_j(E)\, dE \label{pop-eqn}.
\end{equation}
Here ${W_{th}}$ is the threshold for the j$^{th}$ process; $P_j(E)$ is the probability of the j$^{th}$
process at energy E, which can be calculated as $P_j(E)=\sigma_j(E)/\sigma_{in}(E)$; $\sigma_{in}(E)$ is
the total inelastic collision cross section at energy E. 

The numerical yield spectrum, obtained as output of the model,
can be represented in an analytical form as \cite{Bhardwaj99a}
\begin{equation}
      U(E,E_0)=U_a(E,E_0)\ H(E_0-E-E_m)+\delta(E_0-E),  \label{ays-basic}
\end{equation}
where $H$ is the Heavyside function, 
$E_m$ is the minimum threshold of the processes considered, 
and $\delta(E_0-E)$ the Dirac delta function which accounts for the 
collision  at source energy E$_0$. Green et al. \cite{1977Green} have given 
a simple analytical representation for U$_a(E,E{_0})$ as 
\begin{equation}
      U_a(E,E_0)=A_1\xi _0^s+A_2(\xi _0^{1-t}/\epsilon^{3/2 +r}) \label{ays-actual}
\end{equation}
where $\xi=E_0/1000$ and $\epsilon=E/I$ ($I$ is the lowest ionization 
threshold, and $A_1$, $A_2$, t, r, and s are fitting parameters. The fitting parameters 
for xenon gas are $A_1=0.035,\ 
A_2=1.75,\ t=0.0,\ r=-0.065,$ and $s=-0.085$.   

\begin{figure}
\noindent\includegraphics[width=0.75\linewidth]{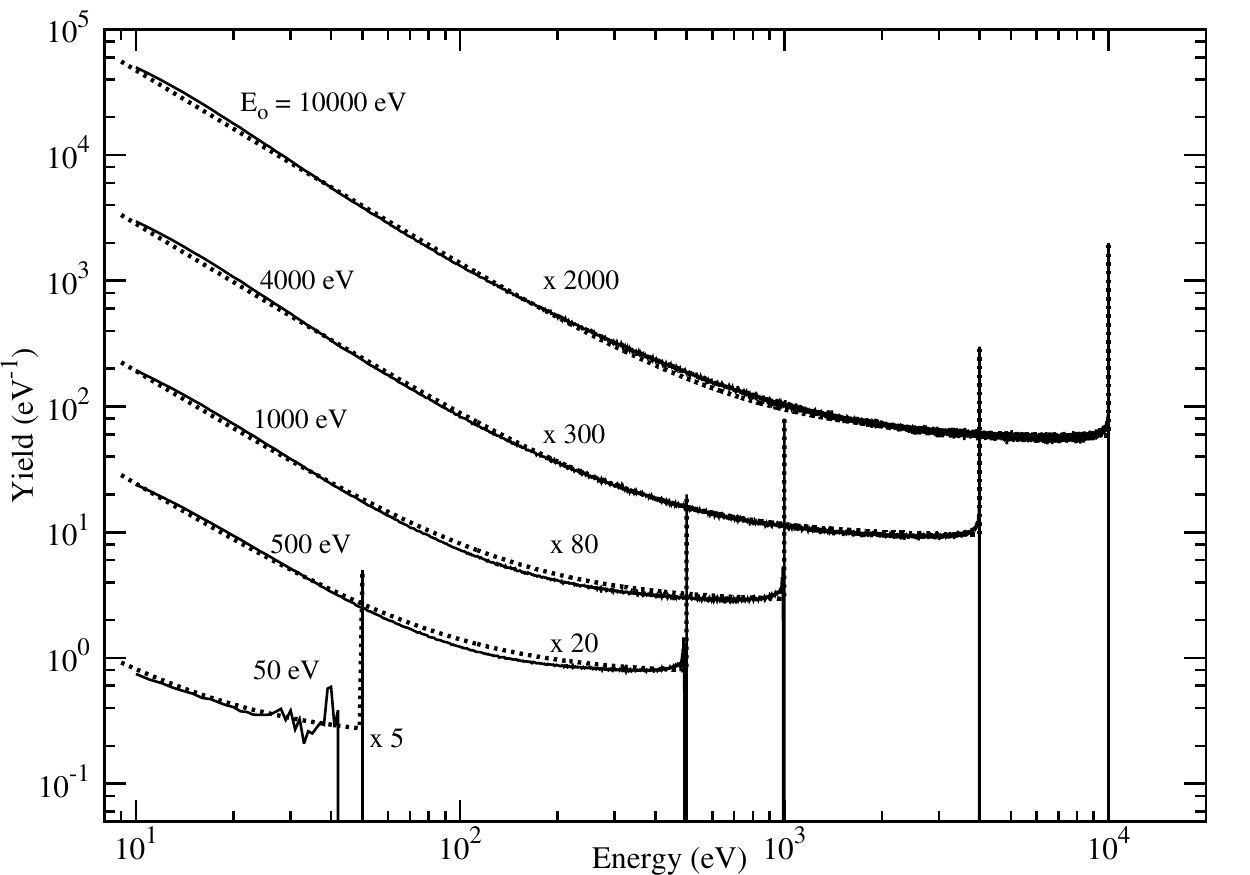}
\caption{Yield spectra for different incident energies. Solid curve shows 
numerical yield spectra obtained using the model. Analytical Yield Spectrum (AYS), 
calculated using equation (\ref{ays-actual}), is represented by dashed curves.
To separate the curves for better clarity, yield spectrum 
for 10000, 4000, 1000, 500 and 50 eV are shown after multiplying with scaling factors 2000, 300, 
80 and 20 and 5, respectively.}
\label{nys&ays}
\end{figure}

Figure \ref{nys&ays} shows numerical yield spectrum (NYS) as well as analytical yield spectrum (AYS) 
for five different incident electron energies. Rapid oscillations seen in the yield spectrum at 
energies close to incident electron energy are not taken into account in our analytical fit.
These oscillations occur due to the fact that energy loss processes are discrete in nature.
For an electron having incident energy E$_0$, an inelastic collision with threshold energy 
E$_m$ will bring the energy down to a value of E$_0$ - E$_m$.  No energy value in the region between 
E$_0$ and E$_0$ - E$_m$ can be acquired by the electron. This is known as Lewis effect \cite{1975Douthat}. 
The heavyside function in  equation (\ref{ays-basic}) accounts for the Lewis effect.

Using the yield spectrum, population of various excitation states can be calculated through
equation (\ref{pop-eqn}). This is useful to determine various properties of gas, like 
mean energy per ion pair and efficiencies of different loss channels which are described 
in the following sections.

\subsection{Mean energy per ion pair}

\begin{figure}
\noindent\includegraphics[width=1.0\linewidth]{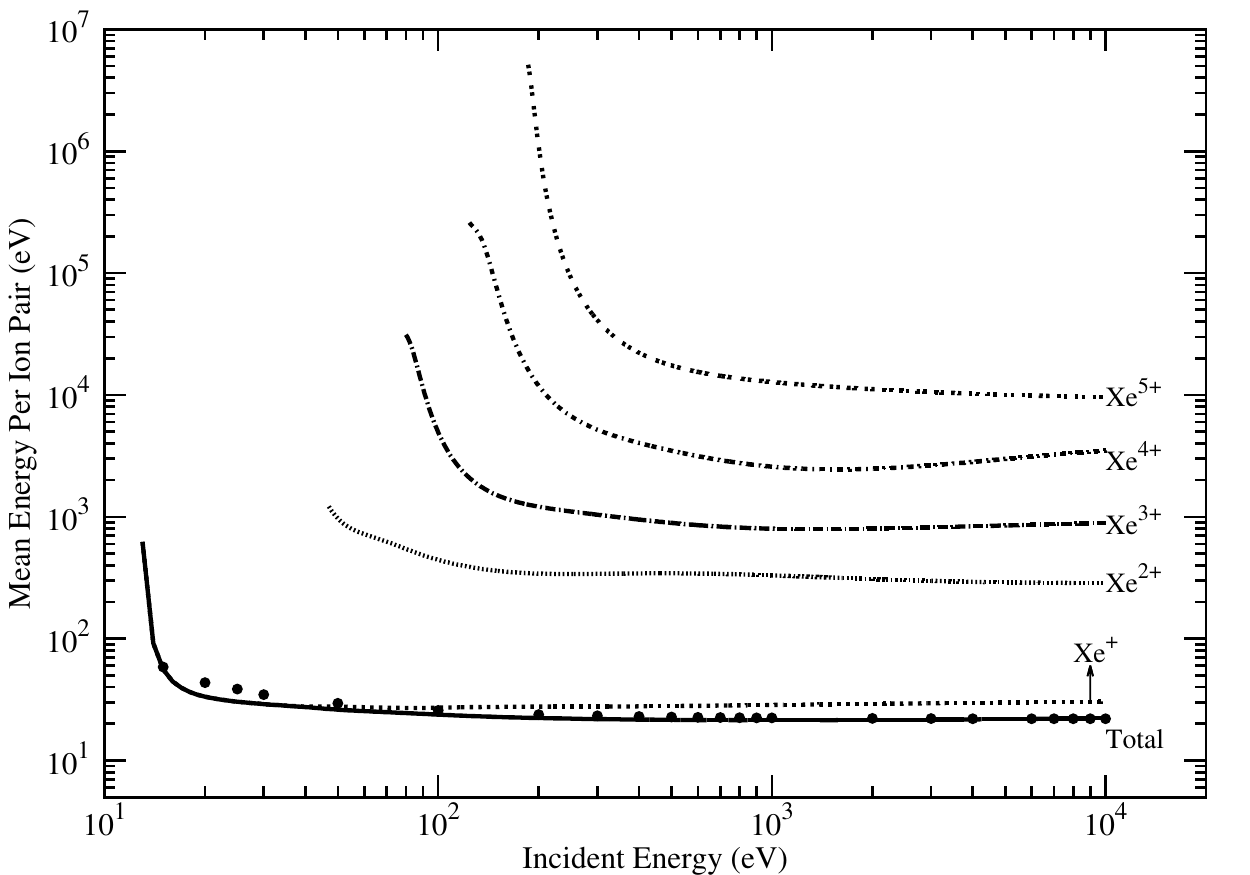}
\caption{Mean energy per ion pair for ions Xe${^+}$, Xe$^{2+}$, 
Xe$^{3+}$,
Xe$^{4+}$, Xe$^{5+}$ and neutral Xe (shown as total). 
Symbols show the values calculated using numerical yield spectrum for Xe}
\label{mepip-figure}
\end{figure}

Mean energy per ion pair, also known as w-value, is the average energy lost by the incident electron in forming an 
electron-ion pair. The w-value for an incident electron energy E$_0$ is calculated as 
\begin{equation}
     w(E_0)=E_0/J(E_0),  \label{mu-eqn}
\end{equation}
where $J(E_0)$ is the population of the ionization events. Figure \ref{mepip-figure} shows the mean energy per ion pair value 
calculated for neutral xenon and for the various ionization channels of Xe. At high 
incident electron energies w approaches a constant value. As the incident electron energy 
decreases, ionization population also decreases since excitation process starts dominating due to their
higher cross section at these energies. Thus w increases  as the incident particle energy decreases.
This behavior of w agrees well with the previous calculations of Combecher\cite{1980Combecher}, Date et al.\cite{2003Date}, 
Dayashankar\cite{1982Dayashankar} and Dias et al.\cite{1993Dias}. 
Mean energy per ion pair calculated for neutral xenon and for the various ionization states 
of Xe at two different incident energies, 10 keV and 300 eV are shown in Table \ref{w-value}.
Date et al. \cite{2003Date} reported a w-value of 21.7 eV at 10 keV. 
Combecher \cite{1980Combecher} measured w-value for electrons in xenon and obtained a value of 22 eV for high 
energy electrons. Dias et al.\cite{1993Dias} obtained a value of 22 eV at 10 keV, while
Dayashankar \cite{1982Dayashankar} calculated a value of 23.1 eV for energy 
$>$200 eV. Our calculated value of mean energy per ion pair is in good agreement with those reported previously.  

\subsection{Secondary electron distribution}

\begin{figure}
\noindent\includegraphics[width=1.0\linewidth]{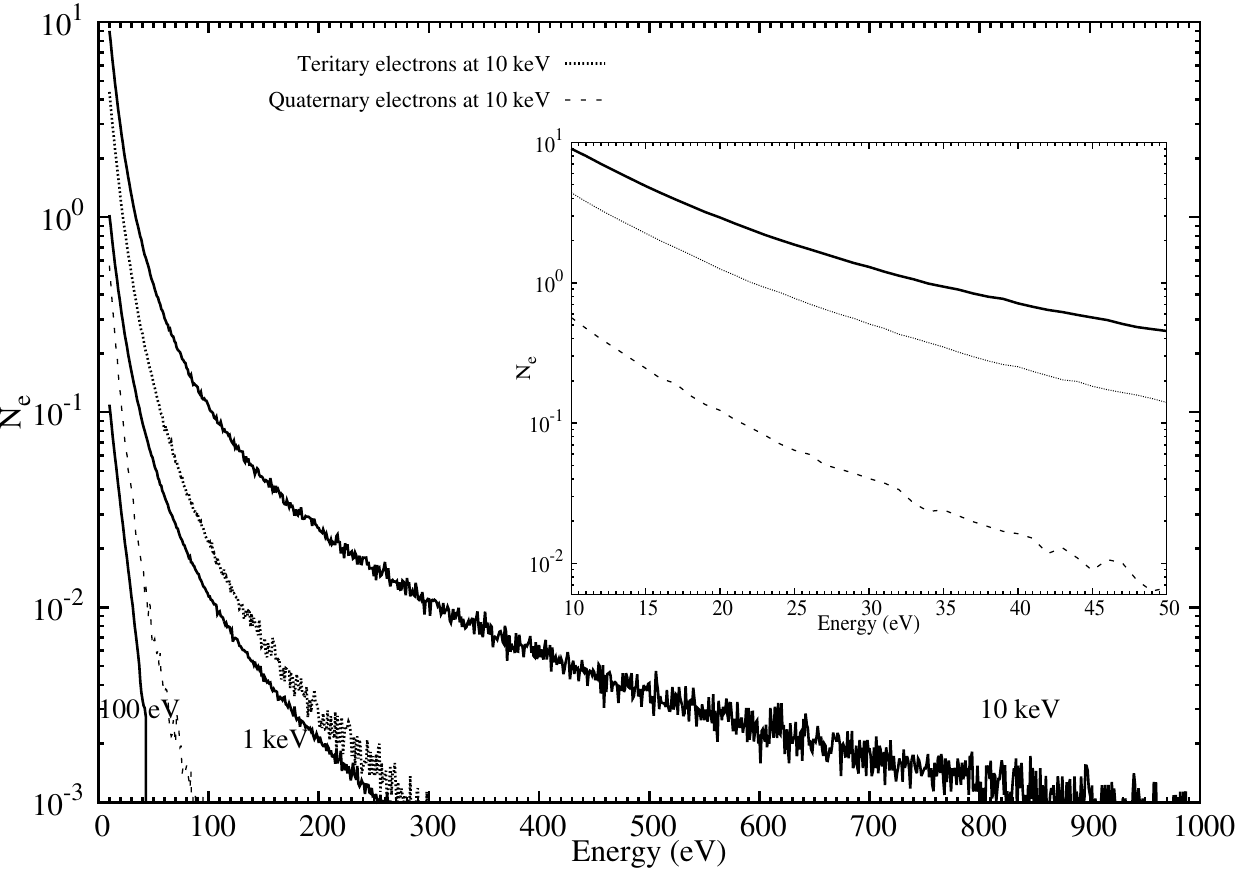}
\caption{Energy distribution of secondary electrons for incident energies 100 eV, 
1 keV and 10 keV.
Y axis shows the number of secondary electrons produced per incident 
primary electron. Dotted and dashed
curves shows distribution of tertiary and quaternary electrons, respectively, 
for an incident energy of 10 keV. The inset shows the energy distribution of secondary, tertiary and 
quaternary electrons for an incident energy of 10 keV by zooming in the lower energy range of 10 to 50 eV.}
\label{2e-dis}
\end{figure}

Secondary electrons generated during ionization can also cause 
inelastic collisions, provided they have sufficient energy. Energies of 
secondary electrons are calculated using equation (\ref{2e-eqn}), and the number of 
secondary, tertiary, quaternary etc electrons is recorded in appropriate 
energy bins. The energy distribution of secondary electrons for different incident 
electron energies is shown in Figure \ref{2e-dis}. Also shown in the same Figure is the distribution 
of tertiary and quaternary electrons for an incident energy of 10 keV. It is clear 
from the figure that during  degradation, an electron with 10 keV energy, will produce 
at least one secondary or tertiary electron whose energy is $<$34 eV, which is still sufficient 
to cause an inelastic collision.

\subsection{Efficiency}

The efficiency of each of the various inelastic processes j can be calculated as 
\begin{equation}
      \eta_j(E_0)=\frac{W_{th}}{E_0}\; J_j(E_0) \label{eqn15} 
\end{equation} where $W_{th}$ is the threshold for the $j^{th}$ process. 

\begin{figure}
\noindent\includegraphics[width=1.0\linewidth]{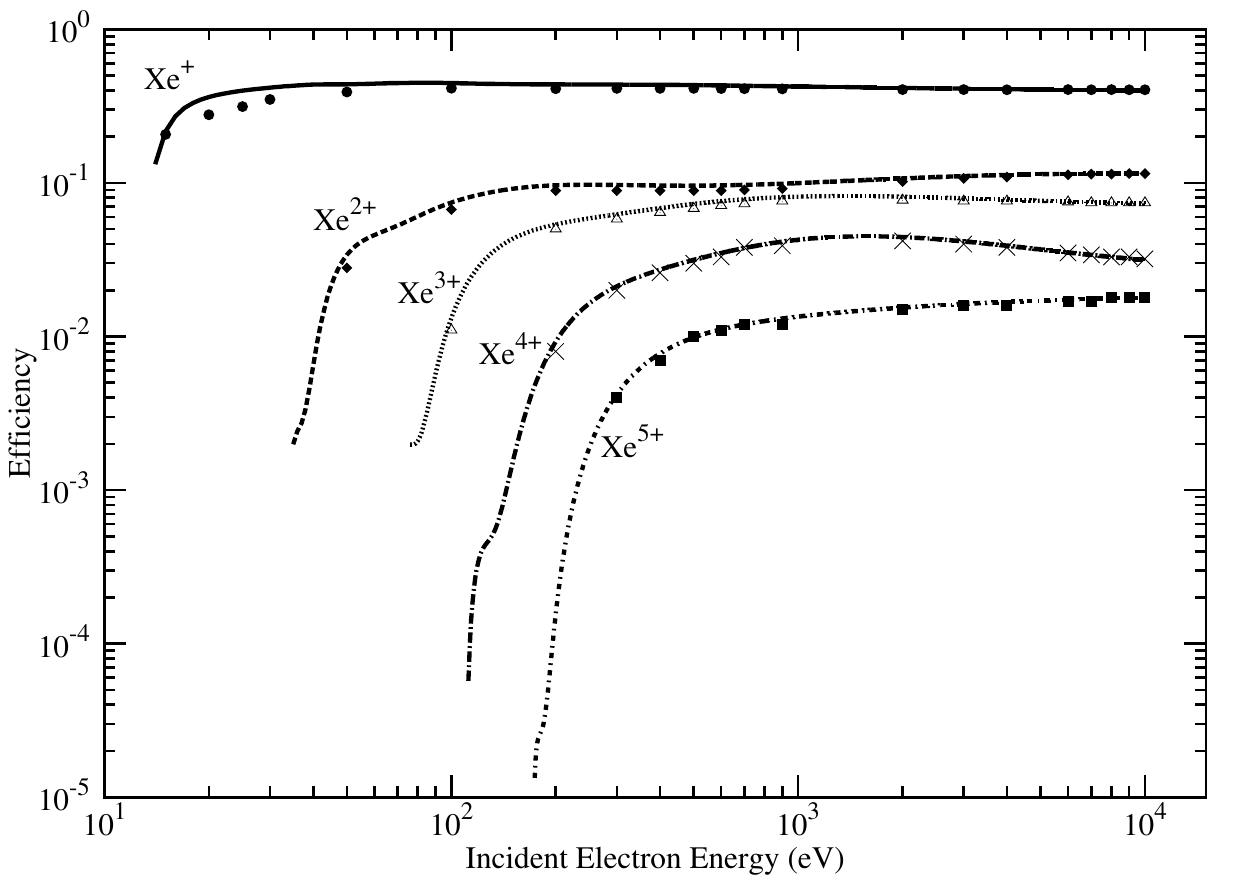}
\caption{Efficiencies of various ionization processes. Symbols
represent the efficiencies calculated using numerical yield spectra 
and solid
lines are efficiencies calculated using AYS.}
\label{ion-e}
\end{figure}

Figure \ref{ion-e} shows the efficiencies of the various ionization channels. 
Efficiencies are calculated using both numerical yield as well as analytical yield and are compared with 
each other. A good match is observed between the values obtained using the two methods. 
Throughout the energy range, Xe$^+$ ionization channel is found to have the maximum 
efficiency due to its high cross section. At 10 keV, Xe$^+$ has an efficiency of 40.5\%.
Xe${^2}{^+}$, Xe${^3}{^+}$, Xe${^4}{^+}$
and Xe${^5}{^+}$ have efficiencies of 11.5\%, 7.4\%, 3.2\% and 1.8\%, respectively. 

\begin{figure}
\noindent\includegraphics[width=1.0\linewidth]{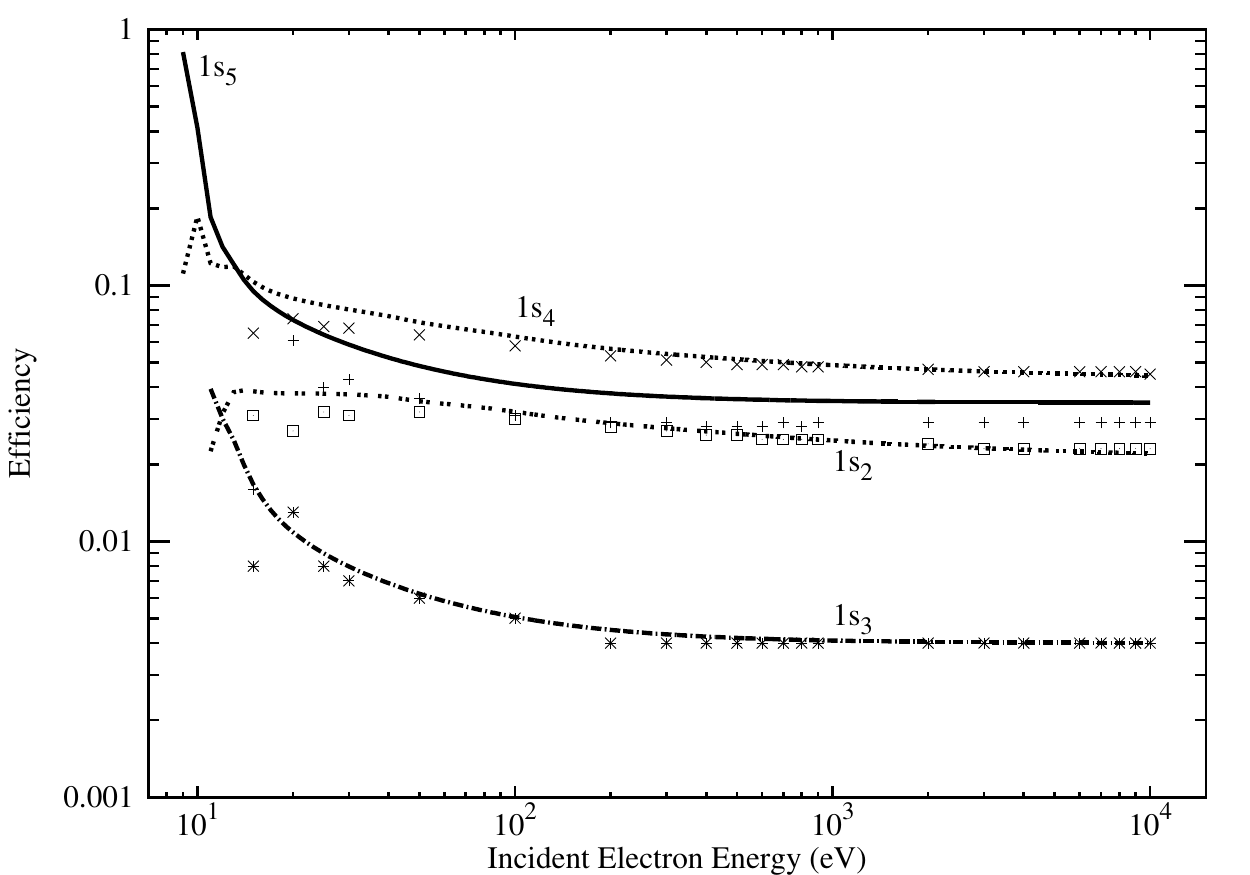}
\caption{Efficiencies of various levels in the 1s configuration. Symbols
represent the efficiencies calculated using numerical yield spectra 
and solid lines are efficiencies calculated using AYS.}
\label{1s-eff}
\end{figure}

\begin{figure}
\noindent\includegraphics[width=1.0\linewidth]{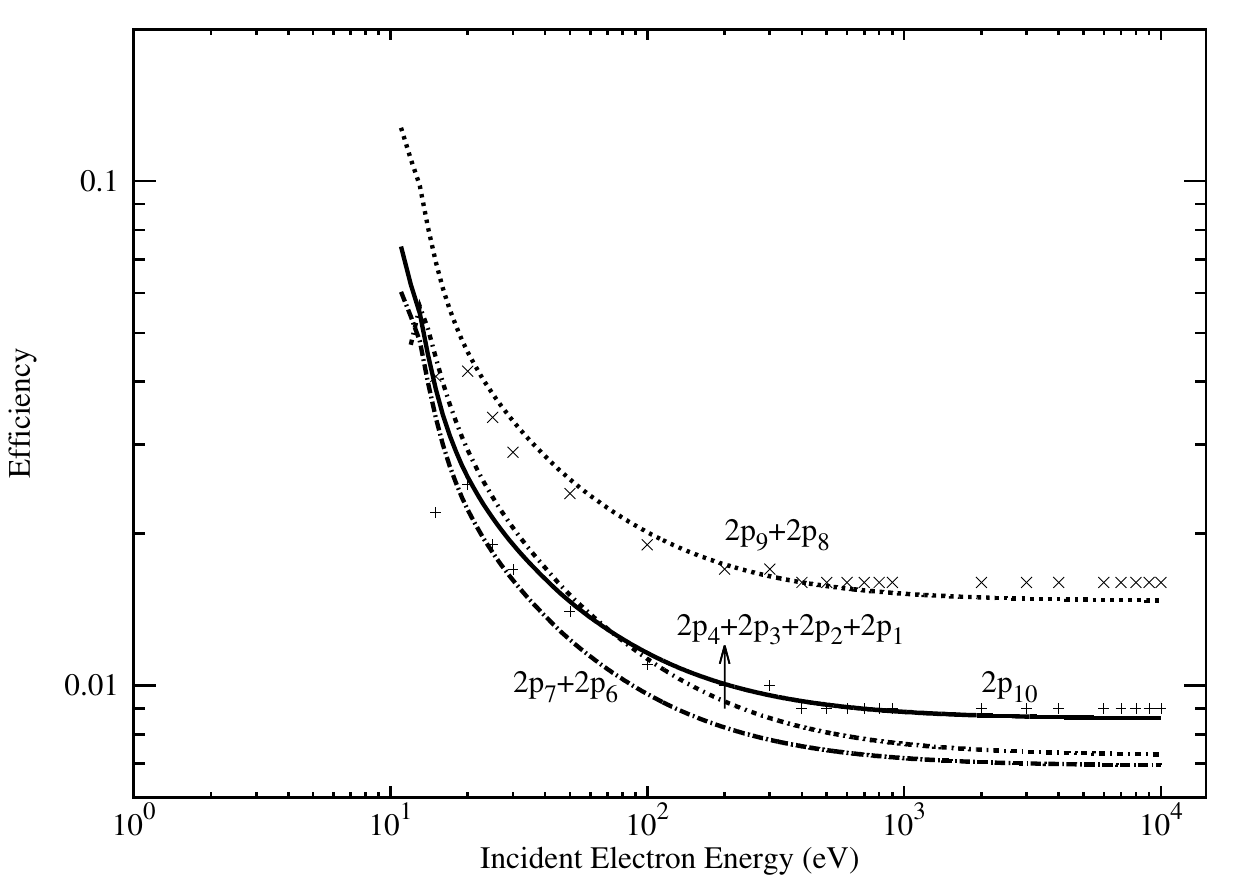}
\caption{Efficiencies of various levels in the 2p configuration. Symbols
represent the efficiencies calculated using numerical yield spectra 
and solid lines are efficiencies calculated using AYS.}
\label{2p-eff}
\end{figure}

\begin{figure}[h]
\noindent\includegraphics[width=1.0\linewidth]{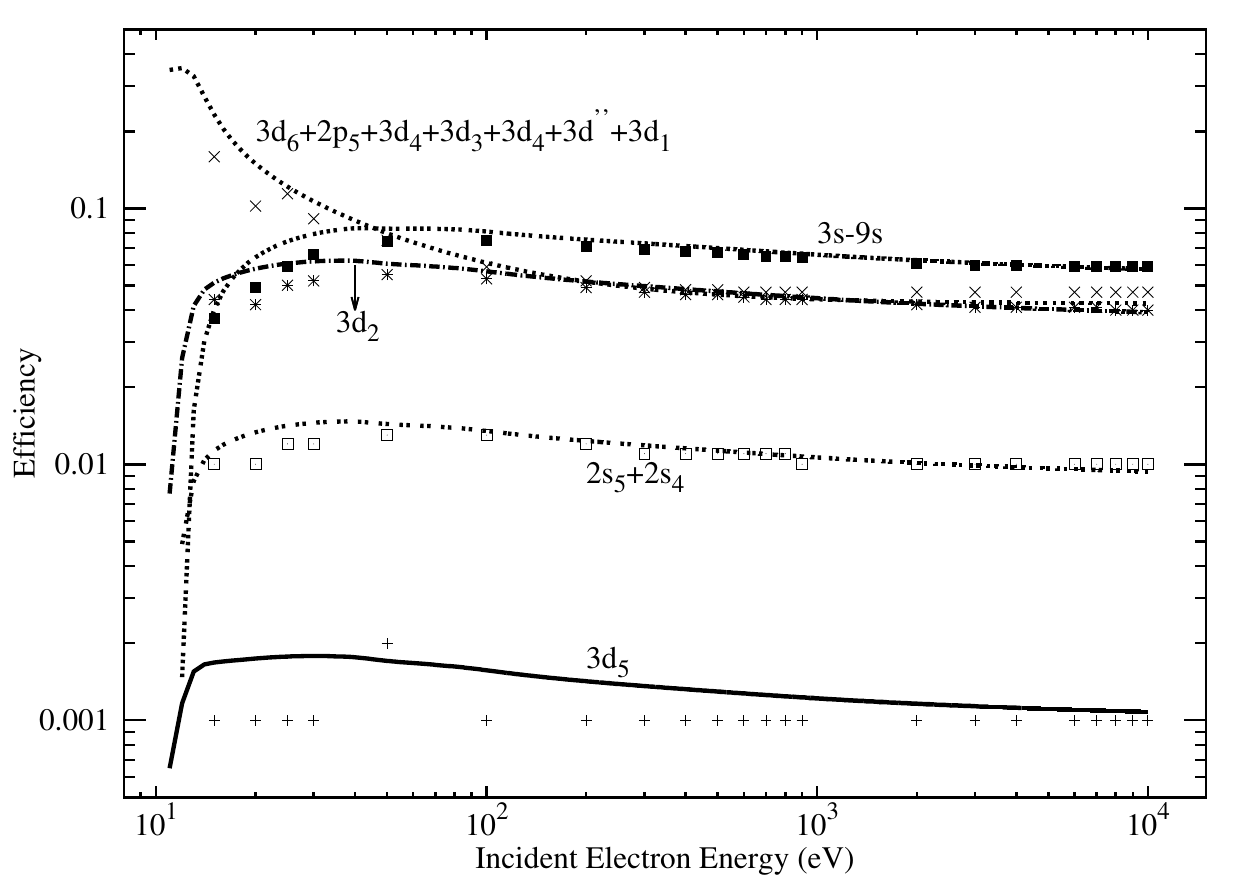}
\caption{Efficiencies of various excitation channels. Symbols
represent the efficiencies calculated using numerical yield spectra 
and solid lines are efficiencies calculated using AYS.}
\label{exct-r}
\end{figure}

Efficiencies of various levels in the 1s configuration are shown in Figure \ref{1s-eff}.
For an incident electron energy of 10 keV, $\sim$10\% of the energy is spent in the 1s configuration.
As seen in the figure, the allowed excitation 1s$_4$ has the highest efficiency throughout 
the energy range with a value of 4.5\% at 10 keV and the lowest efficiency is for the forbidden 
excitation 1s$_3$ with 0.4\% efficiency. The other two excitations 
1s$_2$ and 1s$_5$, have efficiencies 2.3\% and 2.9\%, respectively.
Figure \ref{2p-eff} shows the efficiencies of 2p configuration. The 2p$_9$+2p$_8$ level has an efficiency of 1.6\% at 10 keV
and is the highest among various levels in 2p configuration. Out of $\sim$4\% efficiency of 2p configuration at 10 keV, 
0.9\% is channeled in to 2p${_1}{_0}$ level, 0.8\% into 2p$_7$+2p$_6$ and 0.8\% into 2p$_4$+2p$_3$+ 2p$_2$+2p$_1$ levels.
Efficiencies of remaining excitation channels are shown in Figure \ref{exct-r}. The 3d$_5$ level has a very 
low efficiency of 0.1\% at 10 keV. A combination of various forbidden levels, 3d${_6}$+2p$_5$+3d$_4$+3d$_3$+3d$_4$+3d$^{''}$+3d$_1$, 
has an efficiency of 4.7\%. 
The upper allowed excitation levels (3s-9s) consume around 5.9\% of the incident electron energy.
The 3d${_2}$ and 2s$_5$+2s$_4$ levels have efficiencies of 4\% and 1\% at 10 keV, respectively.
Efficiencies of various inelastic processes at two different incident energies 300 eV and 10 keV are shown in Table \ref{eeta}. 

\begin{figure}[h]
\noindent\includegraphics[width=1.0\linewidth]{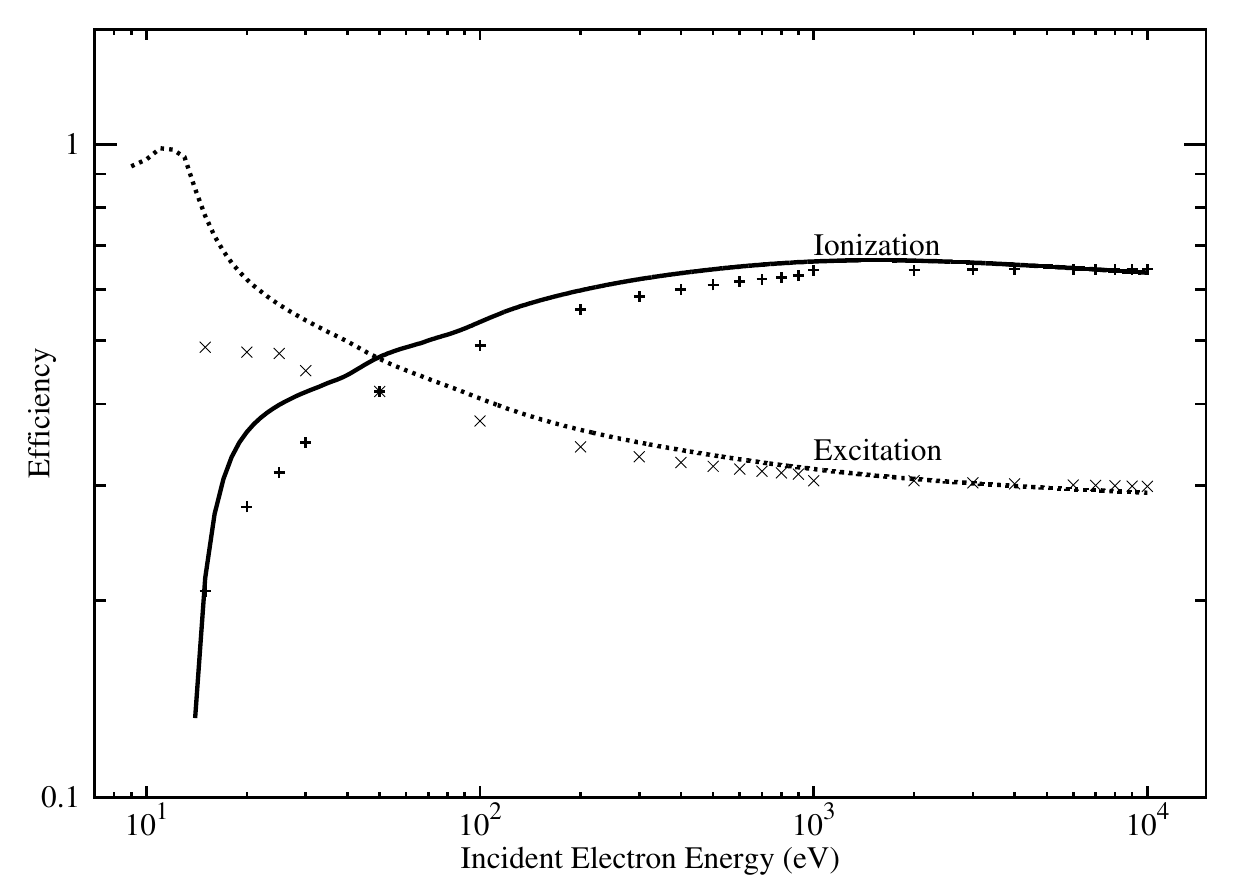}
\caption{Efficiencies of ionization and excitation processes. Symbols
represent the efficiencies calculated using numerical yield spectra 
and solid lines are efficiencies calculated using AYS.}
\label{eff-sum}
\end{figure}

Figure \ref{eff-sum} shows how the incident electron energy is divided among ionization and 
excitation processes. From 50 eV onwards, ionization is the dominant inelastic process. More than 
50\% of incident energy is spent into ionization at these energies. Above 1000 eV, ionization 
efficiency attains a constant value of $\sim$64\%. Excitation dominates at energies less than 30 eV. 
In the energy range were only elatic and excitation collision can occur, 
excitation efficiency is found to be around $\sim$90\% which is consistent with the results of 
 Dias et al.\cite{1993Dias} and Santos et al. \cite{1994Santos}.
At incident electron energies of 10 keV around 30\% of the energy is spent on excitation events.

\section{Dependence of model results on cross sections}

%
To test the dependence of the model results on electron impact cross sections which are used as input to the model, 
we made a test run of the simulation for an incident electron energy of 200 eV. We have run the model by using ionization 
cross section measurements for the Xe$^+$ state by Wetzel et al.\cite{1987Wetzel} which are higher than that of Rejoub et al.
\cite{2002Rejoub} (which we have used in the model) to a maximum of 20\%. The w-value obtained in this case is 23.1 eV 
while ionization efficiency increases from 55\% to 57\% and excitation efficiency 
decreases from 34\% to 32\%. Similarly, when the excitation cross sections for the 1s$_3$ state is replaced
by NIFS recommended cross sections \cite{2003Hayashi}, which is higher than the cross sections of Peuch and Mizzi \cite{1991Puech} (which we have used in the model) 
by a factor of 2, change in the w-value as well as ionization and excitation efficiencies is less than 1\%.

We also tried doubling or halving the major ionization and excitation cross sections to see the impact on model results. The four cases considered are for incident electron energy of 200 eV :\\
Case 1: The ionization cross section for Xe$^+$ state is doubled.\\
Case 2: The ionization cross section for Xe$^+$ state is halved.\\
Case 3: The excitation cross section for 1s$_3$ is doubled.\\
Case 4: The excitation cross section for 1s$_3$ is halved. \\
The w-values and efficiencies obtained for each case is shown in the Table \ref{sensitivity}. 
Doubling or halving the Xe$^+$ cross sections cause a difference of $\sim$5\% in w value and $\sim$3\%
in ionization efficiency. The variation in excitation efficiency in this case is only $\sim$1\%. 
As expected, variation of 1s$_3$ cross sections is not having much effect on w value or efficiencies.

We also observed that, when all the ionization cross sections are doubled keeping excitation cross sections unchanged, 
the w-value shows a decrease of ~12\% (20.7 eV) and ionization efficiency increases to 63\%. When all the excitation cross 
sections are doubled without changing ionization cross sections, the w-value increases by 22\% (29.2 eV) and ionization 
efficiency decreases to 45\% and excitation efficiency increases to 44\%. 
 
\section{Conclusion}
We have developed a Monte Carlo model for degradation of electrons with energy $\leq$10 keV in neutral xenon gas. 
Electron impact cross sections for elastic and various inelastic processes were compiled based on the recent 
experimental and theoretical studies. Analyticaly fitted form of these cross sections are used as input data to the model. 
The numerical yield spectrum calculated using Monte Carlo simulation is analytically represented
through equation \ref{ays-actual} thus generating analytical yield spectrum. 
A good agreement is observed between numerical yield spectrum and AYS.  
From these results the mean energy per ion pair and the efficiency of inelastic processes have been calculated.
The value of mean energy per ion pair is 22 eV for an incident energy of 10 keV which is consistent with the values obtained in 
earlier studies\cite{2003Date,1993Dias,1980Combecher,1982Dayashankar}. 
Secondary electron energy distribution is shown in Figure \ref{2e-dis}. Efficiency calculations showed that 
ionization process dominates for incident energies $>$50 eV and is found to have an efficiency of $\sim$65\% at 10 keV. 
Efficiency of excitation is $\sim$30\% at 10 keV incident energy. Our results are consistent with the previous 
calculations of Santos et al.\cite{1994Santos} and Dias et.al\cite{1993Dias}. 

Results presented in this paper will be useful to understand electron energy degradation process in xenon.
The AYS derived using the Monte Carlo model can be used to calculate steady state electron flux 
in a medium like planetary atmospheres \cite{Bhardwaj99b, 1990Bhardwaj}, ion thrusters etc.  as well as to calculate 
excitation rates or emission intensities \cite{2012Bhardwaj, 2015Jain}. Efficiencies can be used to calculate volume 
production rate  multiplying by electron production rate and integrating over energy. 

\section*{References}

\bibliography{xenonreferences}

\section*{Figures and Tables}

\begin {landscape}
\begin{longtable}{p{0.4in}cccccccccl}
\captionsetup{singlelinecheck=off}
\caption{Elastic differential cross section for xenon in units 
of cm$^2$. Value inside the bracket indicates a 
         linearly extrapolated value. Notation 1E-18 means 1 x 10$^{-18}$ }.\\

\hline
	Angles (degree) & 0 & 10 & 20 & 30 & 40 & 50 & 60 & 70 & 80 & \\
        Energy (eV) & & & & & & & &   \\
        \hline
        
5 & 2.34E-015 & 1.74E-015 & 1.20E-015 & 7.52E-016 & 4.49E-016 & 3.00E-016 & 2.72E-016 & 2.97E-016 & 3.06E-016\\
10 & 3.74E-015 & 2.76E-015 & 1.93E-015 & 1.24E-015 & 7.11E-016 & 3.70E-016 & 1.89E-016 & 1.10E-016 & 8.01E-017\\
100 & 4.34E-015 & 1.11E-015 & 1.28E-016 & 2.37E-017 & 6.94E-017 & 4.34E-017 & 4.87E-018 & 5.59E-018 & 2.29E-017\\
500 & 6.10E-015 & 8.66E-016 & 1.29E-016 & 4.46E-017 & 1.70E-017 & 1.06E-017 & 1.06E-017 & 8.27E-018	& 3.18E-018\\
1000 & 7.07E-015 & 5.71E-016 & 7.86E-017 & 2.49E-017 & 1.25E-017 & 7.52E-018 & 4.55E-018 & 2.82E-018 & 2.36E-018\\
5000 & (9.65E-015) & (1.82E-016) & (2.37E-017) & (8.18E-018) & (4.92E-018) & (1.68E-018) & (6.82E-019) & (1.13E-018) & (5.08E-018)\\
10000 & (1.10E-014) & (1.11E-016) & (1.41E-017) & (5.07E-018) & (3.29E-018) & (8.80E-019) & (3.01E-019) & (7.65E-019) &  (7.07E-018)\\
\hline
\hline
Angles (degree) & 90 & 100 & 110 & 120 & 130 & 140 & 150 & 160 & 170 & 180\\
        \hline
	Energy (eV) & & & & & & & & & &  \\
        \hline

5 & 2.57E-016 & 1.56E-016 & 4.99E-017 & 2.44E-018 & 6.63E-017 & 2.55E-016 & 5.35E-016 & 8.31E-016 & 1.06E-015 & 1.14E-015\\
10 & 6.43E-017 & 5.03E-017 & 4.14E-017 & 4.73E-017 & 7.73E-017 & 1.29E-016 & 1.99E-016 & 2.70E-016 & 3.23E-016 & 3.43E-016\\
100 & 2.38E-017 & 1.20E-017 & 7.70E-018 & 1.16E-017 & 8.96E-018 & 5.31E-019 & 1.06E-017 & 5.58E-017 & 1.15E-016 & 1.42E-016\\
500 & 6.85E-019 & 4.20E-018 & 1.00E-017 & 1.11E-017 & 5.57E-018 & 8.15E-019 & 6.92E-018 & 2.60E-017 & 4.79E-017 & 5.76E-017\\
1000 & 2.82E-018 & 3.11E-018 & 2.33E-018 & 8.21E-019 & 2.47E-019 & 2.54E-018 & 8.32E-018 & 1.61E-017 & 2.29E-017 & 2.55E-017\\
5000 & (1.89E-018) & (2.34E-019) & (2.31E-020) & (7.26E-022) & (1.85E-012) & (1.91E-017) & (2.79E-018) & (1.83E-018) & (1.63E-018) & (1.58E-018)\\
10000 & (1.59E-018) & (7.68E-020) & (3.17E-021) & (3.51E-023) & (1.69E-009) & (4.55E-017) & (1.74E-018) & (7.17E-019) & (5.21E-019) & (4.78E-019)\\
\hline
       \label{tab-EDCS}
\end{longtable}

\end{landscape}

\begin {landscape}
\begin{longtable}{p{0.4in}cccccccccl}
\captionsetup{singlelinecheck=off}
\caption{Fitting parameters for ionization cross sections of xenon. Notation 1E5 means 1 x 10$^{5}$ }\\
        \hline
	 &I (eV) & A & B$_1$ & B$_2$ & B$_3$ & B$_4$ & B$_5$ & B$_6$ & B$_7$ & B$_8$  \\
        \hline
       Xe$^+$ &12.12 & 5.1810E5 & -5.5272E5 & 4.3084E5 & -1.0138E6 & 4.3057E5 & - & -& -& - \\
       Xe$^{2+}$& 33 & 2.0E5 & -1.9897E5 & 1.4518E6 & -2.8675E7 & 2.1198E8 & -7.046E8 & 1.1679E9 & -9.4364E8 & 2.9592E8\\
       Xe$^{3+}$& 65 & 2.2868E5 & -3.2675E5 & 5.4044E6 & -5.7805E7 & 2.5296E8 & -4.9809E8 & 4.4934E8 & -1.5128E8 & - \\ 
       Xe$^{4+}$& 110 & 1.8596E5 & -1.6909E5 & 8.7999E5 & -1.2579E7 & 5.1109E7 & -4.9230E7 & -7.8086E7 & 1.6624E8 & -7.8863E7\\
       Xe$^{5+}$& 172 & 7.1585E4 & -4.8096E4 & 6.5617E6 & 6.5617E6 &-1.5531E7 & 1.4689E7 & -4.9586E6 & - & - \\
\hline
\label{ion-xs-fitting} 
\end{longtable}
\end{landscape}

\begin{table}
\caption{Parameters to calculate excitation cross sections of Xe. For all levels a$_j$ = b$_j$=1 } 
\label{tab-exct-para}
\begin{tabular}{lccl}
\hline
Level & W$_j$(eV) & F$_{oj}$(F$_{oj}$) & c$_j$\\
\hline
1s$_5$ & 8.315 & 53.7 & 2.0 \\
1s$_4$ & 8.437 & 0.26 & 0.0 \\
1s$_3$ & 9.447 & 27.0 & 2.0 \\
1s$_2$ & 9.570 & 0.19 & 0.0 \\
2p$_{10}$ & 9.580 & 2.57 & 1.0 \\
2p$_9$ + 2p$_8$ & 9.706 & 4.89 & 1.0 \\
2p$_7$ + 2p$_6$ & 9.809 & 2.54 & 1.0 \\
3d$_5$ & 9.917 & 0.01 & 0.0 \\
3d$_6$ + 2p$_5$ + 3d$_4$ + 3d$_3$ + 3d$_{4}^{'}$ + 3d$^{''}$ + 3d$_1$ & 10.11 & 20.0 & 1.0\\
3d$_2$ & 10.40 & 0.395 & 0.0 \\
2s$_5$ + 2s$_4$ & 10.59 & 0.097 & 0.0 \\
2p$_4$ + 2p$_3$ + 2p$_2$ + 2p$_1$ & 11.0 & 5.0 & 1.0 \\
(3s-9s)$_{allowed}$ & 11.7 & 0.689 & 0.0 \\
\hline 
\end{tabular}
\end{table}

\begin{table}
\caption{Mean energy per ion pair (w-value) for neutral and 
different ionization states of Xenon for incident energies 300 eV and 10 keV.} 
\label{w-value}
\begin{tabular}{crr}
\hline
& E$_0 $ = 300 eV & E$_0$ = 10 keV\\
\hline
Xe&23.2 eV & 22 eV\\
Xe$^+$   & 29.3 eV & 29.9 eV\\
Xe$^{2+}$& 369.9 eV & 286.6 eV\\
Xe$^{3+}$&1.1 keV & 878.3 eV\\
Xe$^{4+}$&5.4 keV & 3.4 keV\\
Xe$^{5+}$&43 keV& 9.5 keV\\
\hline
\end{tabular}
\end{table}

\begin{table}[ht]
\caption{Efficiencies of various inelastic processes at two different incident energies 300 eV and 10 keV}
\label{eeta}
\begin{tabular}{lcc}
\hline
& E$_0 $= 300 eV & E$_0$ = 10 keV\\
& (\%)& (\%)\\
\hline\
Ionization&61&64.3\\
\hline
Xe$^+$    & 41.3 & 40.4\\
Xe$^{2+}$ & 8.9 & 11.5\\
Xe$^{3+}$ & 6.8 & 7.4\\
Xe$^{4+}$ & 3.0 & 3.2\\
Xe$^{5+}$ & 1.0 & 1.8\\
\hline
Excitation&32.1 &30\\
\hline
1s$_5$ & 2.8 & 2.9 \\
1s$_4$ & 4.9 & 4.5 \\
1s$_3$ & 0.4 & 0.4 \\
1s$_2$ & 2.6 & 2.3   \\
2p$_{10}$ & 0.9 & 0.9 \\
2p$_9$ + 2p$_8$ & 1.6 & 1.6  \\
2p$_7$ + 2p$_6$ & 0.8 & 0.8   \\
3d$_5$ & 0.1  & 0.1  \\
3d$_6$ + 2p$_5$ + 3d$_4$ + 3d$_3$ + 3d$_{4}^{'}$ + 3d$^{''}$ + 3d$_1$ & 4.8 & 4.7  \\
3d$_2$ &  4.6 & 4.0\\
2s$_5$ + 2s$_4$ & 1.1 & 1.0 \\
2p$_4$ + 2p$_3$ + 2p$_2$ + 2p$_1$ & 0.8 & 0.8  \\
(3s-9s)$_{allowed}$ & 6.7 & 5.9  \\
\hline
\end{tabular}
\end{table}

\begin {landscape}
\begin{longtable}{p{0.4in}cccccc}
\captionsetup{singlelinecheck=off}
\caption{Variation of model results with cross sections for an incident electron energy of 200 eV}\\
        \hline
	& Model results & Case 1 & Case 2 & Case 3 & Case 4 \\
	&& Xe$^+$ cross sections doubled & Xe$^{+}$ cross sections halved & 1s$_3$ cross sections doubled & 1s$_3$ cross sections halved \\
        \hline
w-value &23.8 eV & 25 eV & 22.6 eV & 22.9 eV & 23.8 eV\\   
\hline
Total Ionization efficiency & 56\% & 53\%& 59\% & 56\% & 56\%\\
\hline
Total Excitation efficiency & 34\% & 32\% & 33\% & 33\% & 33\%\\
\hline

\label{sensitivity} 
\end{longtable}
\end{landscape}

\end{document}